\documentclass[a4paper]{jpconf}
\usepackage{graphicx}
\begin{document}
\title{First second of leptons}

\author{Dominik~J.~Schwarz$^1$, Glenn~D.~Starkman$^2$, Maik~Stuke$^1$$^{\ast}$}

\address{$^{1}$Fakult\"at f\"ur Physik, Universit\"at Bielefeld, Postfach 100131, 33501 Bielefeld, Germany, and\\
					$^{2}$ CERCA/ISO, Department of Physics, Case Western Reserve University, Cleveland,OH 44106-7079 }
\eads{\mailto{dschwarz} and \mailto{mstuke} at \mailto{physik.uni-bielefeld.de}, and\\ \mailto{glenn.starkman} at \mailto{case.edu}}

\address{$\ast$: Speaker}

\begin{abstract}
A poorly constrained parameter in the Standard Model of Cosmology is the lepton asymmetry $l = \sum_{f} l_f=\sum_f({n_f+n_{\nu_f}})/s$. Each flavour asymmetry $l_f$ with $f=e$, ${\mu}$, ${\tau}$ is the sum of the net particle density of the charged leptons $n_f$ and their corresponding neutrinos, normalized with the entropy density $s$. Constraints on $l_f$ from BBN and CMB allow for lepton flavour asymmetries orders of magnitudes larger than the baryon asymmetry $b\simeq 10^{-10}$. In this article we show how large lepton (flavour) asymmetries $l_f \leq{\cal O}(0.1)$ influence the early universe, in particular the freeze out of WIMPs and the cosmic QCD transition.
\end{abstract}

\section{Introduction}
The recently reported higher primordial abundance of $^4$He \cite{Izotov} and the larger number of effective number of neutrinos by observations of the cosmic microwave background \cite{WMAP7, ACT} might be a motivation for large neutrino chemical potentials or equivalently large neutrino asymmetries. Charge conservation and neutrality ensures  that the asymmetry in the charged leptons is of the same order as $b=n_b/s(T)\simeq {\cal O}(10^{-10})$ \cite{WMAP7}. But due to the difficulties in detecting neutrinos one can only obtain constraints on the lepton (flavour) asymmetry  from indirect observations.  A large $l_f$ 
would affect big bang nucleosythesis (BBN) in two ways. It changes the primordial $^4$He abundance \cite{Steigman} and the expansion of the universe \cite{Olive}. We show in this article some results on how large lepton (flavour) asymmetries influence the early universe between the electroweak phase transition and the onset of neutrino oscillations. For a more detailed discussion see \cite{Schwarz:2009ii} and \cite{SSS}. 

\section{Pre-BBN}
Before BBN, at $T_{\rm{osc}}\simeq$ few MeV, neutrino flavour starts to oscillate and equilibrates any initial 
flavour asymmetries before the onset of BBN \cite{Nuosc}. For temperatures larger then $T_{\rm{osc}}$ lepton flavour number is conserved. Within the Standard Model, no violation of baryon and lepton number are possible after the electroweak phase transition at $T_{\rm{ew}}\simeq 200$ GeV \cite{PT}. A lepton flavour asymmetry $l_f$ remains constant in the early universe between $T_{\rm{ew}}$ and $T_{\rm{osc}}$. Here we assume $l_f\leq {\cal O}(0.1)$ \cite{Steigman,Schwarz:2009ii}. 

All particle interactions between these two events are in statistical equilibrium and can be described by the set of 5 equations:
\begin{eqnarray}
\label{lf}l_f~s(T)=(n_f+n_{\nu_f})\ \ &&{\rm{for}}\  f=\rm{e},~\mu,~\tau \\
b~s(T)=\sum_i B_i n_b \ \ &&{\rm{with}}\ B_i={\rm{baryon~ number~ of~ species~}} i\\
\label{charge}q ~ s(T)=0=\sum_i Q_i n_q \ \ &&{\rm{with}}~ Q_i={\rm{electric~ charge~ of~ species~}}i.
\end{eqnarray} 
The net particle density of a species $i$ with mass $m_i$ and chemical potential $\mu_i$ is given by 
\begin{equation}
\label{ni}n_i=\frac{g_i}{2\pi^2}\int^{\infty}_{m_i}{E\sqrt{E^2-{m_i}^2}\left(\frac{1}{\rm{exp}\frac{E-\mu_i}{T}\pm1}-\frac{1}{\rm{exp}\frac{E+\mu_i}{T}\pm1} \right)}dE
\end{equation}
with the internal degrees of freedom of a species $g_i$.  Anti-particles $\bar{i}$ have the same chemical potential as the corresponding particle, but inverted sign: $\mu_i=-\mu_{\bar{i}}$. With the observed baryon asymmetry and the charge neutrality of the universe we can solve equations (\ref{lf}) to (\ref{charge}) for different $l_f$ for all Standard Model particles in statistical equilibrium, leaving $\mu_i$ as free parameters. 

\begin{figure}[b]
	\centering
		\includegraphics[width=0.48\textwidth]{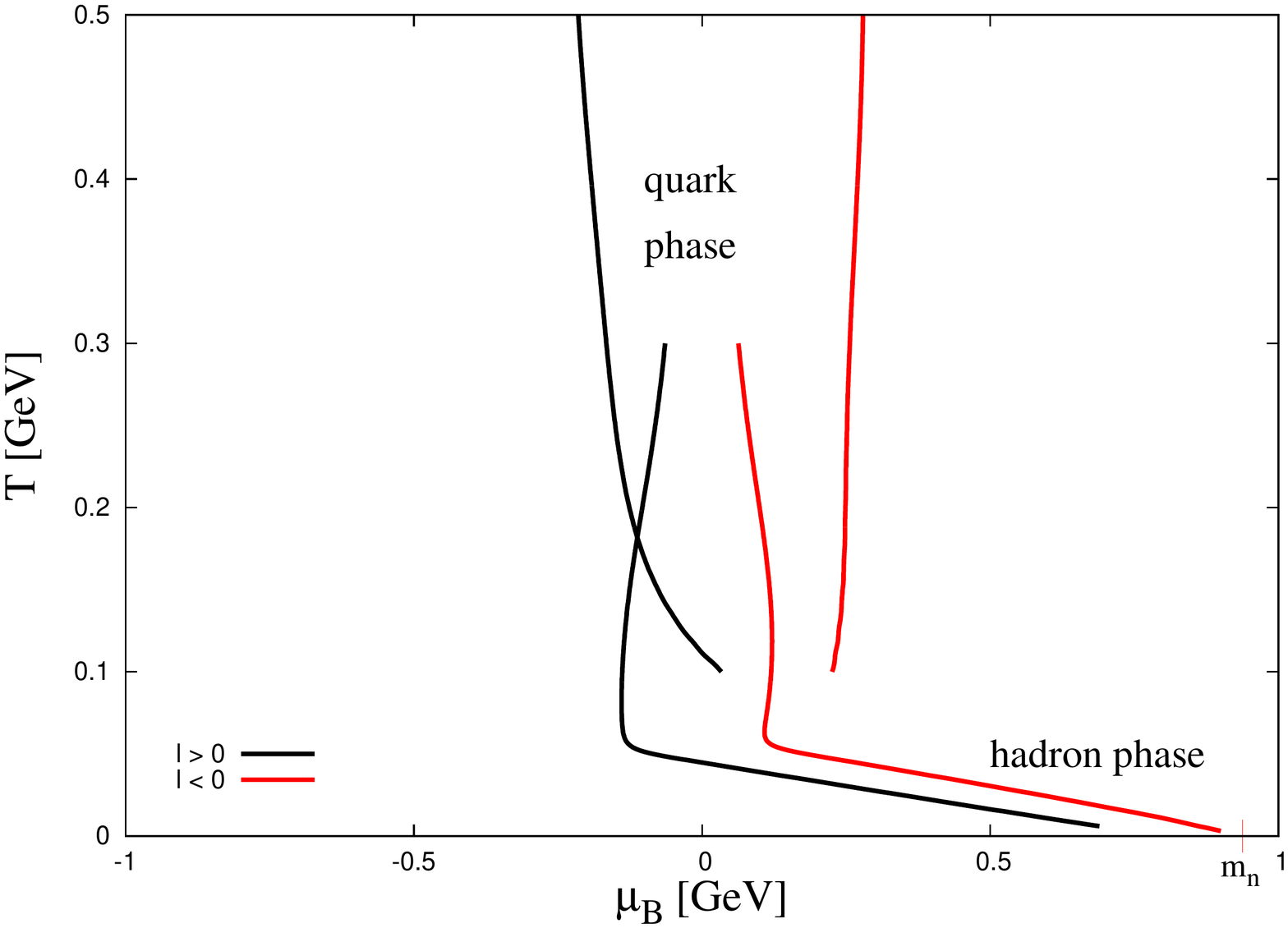}
		\hspace{.1cm}
		\includegraphics[width=0.48\textwidth]{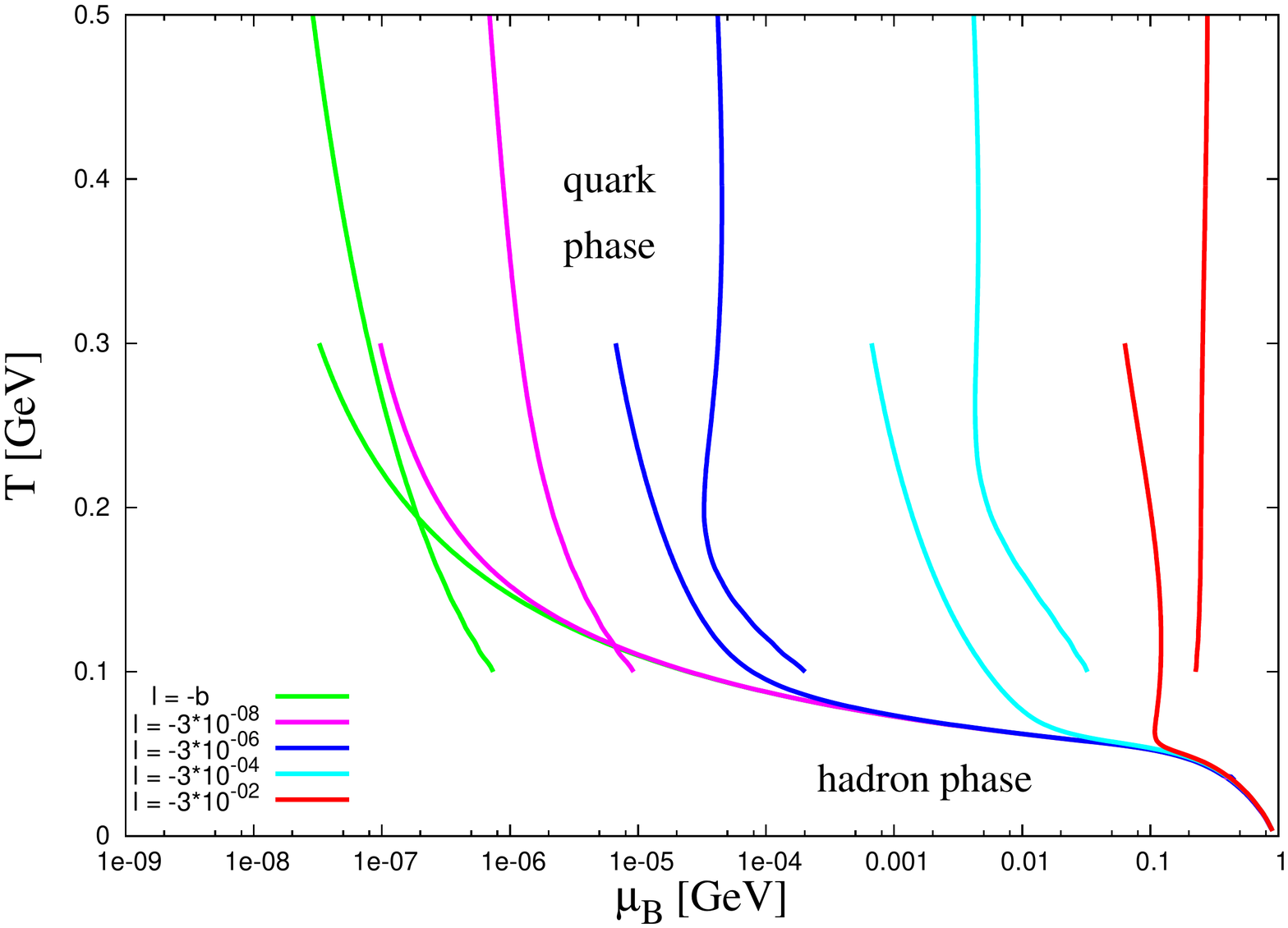}
	\caption{Dependence of the $\mu_B$-T trajectory on sign and size of a lepton asymmetry}
	\label{fig:QCD}
\end{figure}

With each conserved quantum number one can associate a chemical potential which can be calculated from the free energy of the particle plasma. Apart from the three lepton chemical potentials and the charge chemical potential we can also calculate the baryon chemical potential $\mu_b$. The latter one is for cosmologist and particle physicist of greater interest to describe the QCD transition, the confinement of quarks to hadrons. It is still an open question, if the cosmic QCD transition is just a rapid change or a real phase transition. One would like to have deeper knowledge of the order of the transition, since it sets the initial conditions for BBN and the different orders would lead to different observable consequences, like a modification of the primordial garavitational background or quark nuggets \cite{cosmicQCD}. For the description of the transition in the ($\mu_b$--$T$) phase diagram, we found that the order of $\mu_b$ is set for large lepton asymmetries by $l$. Assuming all particles massless and $T\gg T_{\rm{QCD}}$ we find for lepton asymmetries $3l_f=l$ the approximate solution $\mu_b=\left(\frac{39}{4} b  -  l\right) \frac{s(T)}{4 T^2}$. The numerical results for $\mu_b$ including all Standard Model particles at all temperatures with their physical masses are shown in figure \ref{fig:QCD}. We find that sign and size of the lepton asymmetry plays an important role for the description of the cosmic QCD transition in the ($\mu_b$--$T$)-plane. For large asymmetries $l\leq10^{-2}$ the trajectories are pushed in the vicinity of a possible first order transition, predicted by results from lattice simulations and perturbatve calculation of QCD combined with results from heavy ion collisions \cite{Stephanov:2007fk}. It is a good approximation to describe the early universe and even the phase transitions as equilibrium processes \cite{cosmicQCD}, but for a more concrete statement at $T=T_{\rm{QCD}}$ we would have to incorporate the strong interaction.  
To describe the cosmic QCD transition in a universe with baryon, lepton flavour and charge conservation, a two dimensional ($\mu_b$--$T$) phase diagram is not sufficient. More realistic is a five dimensional phase diagram ($\mu_{l_f}$, $\mu_b$, $T$) for a charge neutral and homogeneous universe. For a more detailed discussion see \cite{Schwarz:2009ii}.

Large chemical potentials influence also the energy density and therefore the Hubble rate $H=\sqrt{(8/3)\pi G_{N}}~\sqrt{\epsilon (T,\mu_i)}$, with Newtons constant $G_N$ and the energy density of all particle species (particle and antiparticle) $\epsilon = \sum_{j = i,\bar{i}} {g_j}/({2\pi^2})~\int^{\infty}_{m_j}{E^2(E^2-m_j^2)^{1/2}f_j(E){\rm d}E}$. The distribution function $f_i (E) = ({\rm exp}\frac{E-\mu_i}{T} \pm 1)^{-1}$ describes the boson or fermion distribution for a particle $i$.

\begin{figure}[htbp]
\centering
	\includegraphics[width=0.50\textwidth]{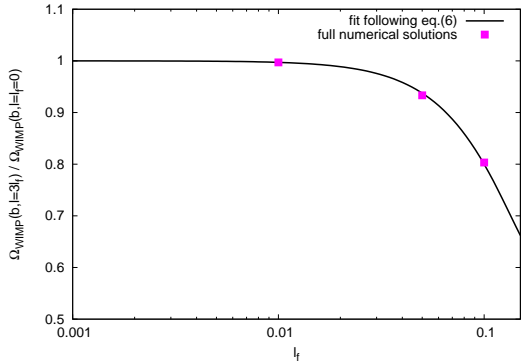}
	\caption{Comparison between the relative WIMP dark matter abundance without lepton flavour asymmetries $\Omega_{DM}(l=l_f=0)$ and large asymmetries $l_f=l_e=l_{\mu}=l_{\tau}$. We observe for $l_f=0.1$ an effect of 20 percent. The full numerical solutions include all Standard Model particles.}
	\label{fig:Deltamulf}
\end{figure}

This influences another major event of cosmology: The freeze out of the best motivated dark matter candidate, the WIMP \cite{Jungman:1995df} with a typical mass of $10$ GeV $<m_{\chi}<1$ TeV. It decouples statistically from the Standard Model particles\footnote{Also called \emph{chemical} decoupling. The WIMPs remain in kinetic equilibrium with the Standard Model particles until $T\simeq {\cal O}$(few MeV).} at $T_{\rm{fo}}=m_{\chi}/25$, when its annihilation rate drops below the Hubble rate. The number density remains constant $\left({n_{\chi}}{s^{-1}} \right)_{\rm today}=\left({n_{\chi}}{s^{-1}} \right)_{\rm{fo}}$ and thus the relative relic abundance $\Omega_{\rm WIMP} {\rm{h}^2} = m_{\chi}n_{\chi}\rho_c^{-1}$ is observable today. The number density follows the Boltzman equation $\dot{n_{\chi}} + 3Hn_{\chi} = -\left\langle \sigma |v| \right\rangle (n_{\chi}^2-n_{\chi,~\rm eq}^2)$ with the thermal average of the cross section, which can be approximated as $\langle \sigma v \rangle= a + b/x + {\cal O}(x^{-2})$. The numbers $a$ and $b$ describe s- and p-wave annihilation and $x \equiv m_{\chi}/T$, for more details see e.g.~\cite{Jungman:1995df}. 
Increasing the energy density via large chemical potentials changes the Hubble rate and thus the temperature, where the freeze out happens. We illustrate this effect by the change of the helicity degrees of freedom in the energy density $g_{\ast}(T,\{\mu_i\})=\frac{30}{\pi^2T^4} \epsilon(T,\{\mu_i\})$ and in the entropy $s(T)=\frac{2\pi^2}{45}T^3g_{S\ast}=-\int{\left[f {\rm ln}f \mp(1 \pm f){\rm ln}(1 \pm f)\right]\frac{{\rm d}^3p}{(2\pi)^3}}$, where upper and lower signs refer to bosons and fermions, respectively. 
The present relative relic abundance becomes then $\Omega_{\rm WIMP}h^2\simeq \frac{8.5 \times 10^{-11}}{\rm{GeV}^2} \left(\frac{\sqrt{g_\ast}}{g_{S\ast}}\right)_{\rm fo} 
\frac{x_{\rm{fo}}}{a + b/x_{\rm{fo}}}$. As long as the deviation $\Delta g{\ast}$ is much smaller than the standard $g_{\ast}(l = l_f=0)$ we can approximate  
\begin{equation}
\label{Omega}\frac{\Delta \Omega_{\rm WIMP}}{\Omega_{\rm WIMP}} = 
\frac 12 \left( 1 - (1 + \ell) \frac 1{x_{\rm fo}} \right) 
\frac{\Delta g_\ast}{g_\ast} - \frac{\Delta g_{S\ast}}{g_{S\ast}},
\end{equation} 
with $\ell = 0,1$ depending on the s- or p-wave annihilation channel domination. For the numerical analysis we assume the governing effect in helicity degrees of freedom and neglect the effect in $x_{\rm{fo}}\propto c~\ln g_{\ast}$. This leads to:
\begin{equation}
\frac{\Omega_{\rm WIMP}(l ,l_f\neq0)}{\Omega_{\rm WIMP}(l,l_f=0)} \sim \left[\sqrt{\frac{g_{\ast}(l,l_f\neq0)}{g_{\ast}(l,l_f=0)}}~ \frac{g_{S\ast}(l,l_f=0)}{g_{S\ast}(l,l_f\neq0)}\right]_{x_{\rm{fo}}}
\end{equation}            
A numerical analysis with all physical particle masses and not restricted to small chemical potentials 
is shown in figure \ref{fig:Deltamulf} for different lepton asymmetries $l_f=\frac{1}{3}l$. We show the ratio $\Omega_{\rm{WIMP}}(l,l_f)/\Omega_{\rm{WIMP}}(l=l_f=0)$ as a function of lepton flavour asymmetry $l_f$ and observe an effect of order 1 $\%$ for $l_f=0.01$ and of almost $20\%$ for $l_f=0.1$.

\section{Conclusion}
In this article we have shown that  large lepton asymmetries do have an impact on the evolution of the early universe. We focused on the cosmic QCD transition and the WIMP freeze out.

We have shown that $b\ll |l|\leq 0.01$ can significantly influence the dynamics of the QCD phase transition and maybe trigger even the order of the transition in the $(\mu_b-T)$-plane.
However, a sufficient description of the cosmic QCD transition needs at least 5 dimensions ($l_f$, $b$,T, $q=0$) and even more for an inhomogeneous universe.

We have further shown that large lepton flavour asymmetries can decrease the relative relic abundance of WIMP dark matter for a given cross section. For symmetric flavour asymmetries $l_e=l_{\mu}=l_{\tau}$ and $l_f\geq{\cal O}(0.01)$ we found a few percent effect on the relic WIMP abundance. This might allow for smaller cross sections for a WIMP candidate to match the observed dark matter abundance.
\section*{Acknowledgements}
The speaker thanks the organizers for the inspiring meeting. This work was supported by the Friedrich-Ebert-Foundation (M.S) and Deutsche Forschungsgemeinschaft (M.S. and D.J.S.) and by a grant from the US-DOE to the particle astrophysics theory group at CWRU.

\end{document}